\documentclass[onecolumn,aps,prd,preprintnumbers,amsmath,amssymb,showkeys]{revtex4-2}

\usepackage{natbib}
\usepackage{graphicx}
\usepackage{bm}
\usepackage{color}

\usepackage{amsmath}
\usepackage{amsfonts}
\usepackage{amssymb}
\usepackage{makeidx}
\usepackage{graphicx}
\usepackage{physics}
\usepackage[rightcaption]{sidecap}
\usepackage{graphicx}
\usepackage{subcaption}
\begin{document}
	
	\title{{Analysis of memory effects in the dynamic evolution of the spin-boson model}}
	\author{Rayees A Mala$^1$, Mehboob Rashid$^2$, Muzaffar Qadir Lone$^1$\footnote{corresponding author: lone.muzaffar@uok.edu.in} }	
	\affiliation{$^1$ Quantum Dynamics Lab, Department of Physics, University of Kashmir, Srinagar-190006 India,\\
  $^2$Department of Physics, National Institute of Technology, Srinagar-190006 India}

	\begin{abstract}
	Quantum information processing relies on how dynamics unfold in open quantum systems. In this work, we study the 
		non-Markovian dynamics in the single mode spin-boson model at strong  couplings. In order to apply perturbation theory, we transform our 
		Hamiltonian to polaron frame, so that the effective system-bath coupling gets reduced. We employ  coherence defined by $l_1$-norm to analyze the non-Markovian effects  in the spin-boson model. In the transformed frame of reference, the correlation timescales for the bath are significantly shorter than the system’s relaxation timescale—a key assumption for Markovian dynamics. However, intriguingly, we demonstrate that under the large polaron theory, the reduced dynamics exhibit effective non-Markovian behaviour within a specific range of couplings, while remaining Markovian beyond this range.
		

	\end{abstract}
	\maketitle

	\section{Introduction}

The field of quantum mechanics is fundamentally dependent upon the intricate interplay between a well-defined quantum system and its surrounding environment, known as a bath. These open quantum systems represent a novel theoretical framework that holds particular significance for the rapidly growing fields of quantum information and computation  \cite{NC}. 
Specifically, two distinct kinds of open quantum systems can be defined: Markovian  and non-Markovian dynamics\cite{BP,BP1,IND}.
 In a Markovian process, information exclusively propagates unidirectionally—from the system to the bath. This simplified picture, which often occurs in weakly coupled systems, is well described by the Lindblad equations or CPTP maps \cite{DC,LIND,GKLS}. However, as the coupling becomes stronger, the Markovian approximation breaks down, allowing memory effects to occur  \cite{abc,abc1,ADB}. Non-Markovian dynamics, characterized by information flowing back into  system from the bath, differ sharply from their Markovian counterparts. This retrograde flow interrupts the system's temporal evolution, unlike the case in Markovian processes \cite{DC}.
Interestingly, non-Markovian dynamics have proven valuable in various quantum information protocols, such as secure communication (quantum key distribution)  \cite{QKD1,QKD2}, precise measurements (quantum metrology)  \cite{QM1,QM2,QM3}, and teleportation \cite{Mehb1,TNM1,TNM2}. Even the field of quantum biology is thought to harness these memory effects \cite{QB}.

To unveil the essence of non-Markovianity, several quantitative measures have been devised. These measures, often focusing on information flow or deviations from the semi-group property (as detailed in \cite{Rivas1, Benn, BR}), quantify the departure from the ideal Markovian behaviour. A prominent example is the BLP measure, introduced by Breuer et al. \cite{BLP}. It delves into the information exchange between the system and the bath, employing metrics like trace distance to assess the distinguishability of evolving quantum states. Similarly, the RHP measure, developed by Rivas et al. \cite{Rivas2}, captures non-Markovianity through the non-divisibility of the dynamical maps governing the system's evolution. The quest to quantify non-Markovianity extends beyond these measures, encompassing approaches based on quantum Fisher information flow [QFI], coherence measures \cite{coh1, coh2}, and quantum interferometric power \cite{int, int1}. Notably, all these measures share a common thread – the non-divisibility of CPTP maps. This work leverages a specific non-Markovianity measure based on the non-monotonic behaviour of coherence, quantified by the $l_1$-norm\cite{col1}. We then employ this measure to scrutinize the dynamics of the spin-boson model under the influence of strong coupling.

	Spin boson model \cite{Leg,weiss} is a paradigmatic model of a dissipative quantum system describing wide range of physical phenomena. For example, the impurity problems described by Kondo physics \cite{kondo} are often mapped to spin boson model to understand nature of phase transitions and dynamics involved. The transport phenomena in photosynthesis \cite{tra}, thermodynamics \cite{pt} and various light-matter systems \cite{LM} are described by such model. This model has lead to the development of various analytical and numerical techniques for strongly correlated systems \cite{weiss} . In this paper, we consider a version of spin boson model, where the qubit (spin) is strongly coupled to a single bosonic mode \cite{sm,sm1,sm3} instead of usual multimode bath. Such single mode approximation can be related to the continuous quantum measurement \cite{sm} where macroscopic oscillations of a single oscillator are measured by some linear detector modeled by a bosonic mode. There have been intensive research along the direction of non-Markovianity in spin boson model \cite{nms1,nms2,nms3}. 
	We consider the following problem in this work. It is usually believed  \cite{amy,ms} that if the relaxation time scale of the system is larger than the correlation time scale of the bath, then the dynamics is Markovian. In the large polaron theory, the bandwidth or tunneling rate gets reduced, thus making the relaxation time scale smaller in comparison to bath correlation time scale. This implies Markovian dynamics in the polaron frame. However, we re-examine this assumption in the finite time evolution of the reduced system in the polaron frame. We show that a naive application of Markovian approximation of the master equation yields some inconsistent results.

	We organize this work in the following way. The notion of coherence based on $l_1$-norm is introduced in section II. Based on $l_1-$norm, we introduce the concept of non-Markovianity. In section III, the  model system i.e. single mode spin boson system is introduced. In the same section, we perform Lang-Firsov transformation to obtain the effective Hamiltonian in the polaron frame. The dynamical analysis of the model in done in section IV. Finally, we conclude in section V.

	\section{ Coherence and  Non-Markovianity}

	Quantum superposition principle gives rise to notions of coherence and entanglement in quantum systems, that plays crucial role in the foundational understanding of quantum mechanics and its applications to quantum technologies \cite{QC1,QC2,QC3,QC4,QC5}. Many measures of coherence have been proposed like $l_1$-norm, relative entropy, skew information and so on. Following the footprints of quantum resource theory, a measure of quantum coherence must satisfy the following axioms \cite{QC4}:
	
Consider a Hilbert space$\mathcal{H}$ of finite dimension $d$. Let $\mathcal{I}$ denote the collection of incoherent states, characterized by their diagonal representation within a specified basis: $ \rho_I = \sum_{i=1}^{d} a_i |i\rangle\langle i|$, where $ a_i \in [0,1] $ and $ \sum_{i=1}^{d} a_i = 1 $. Here, $ |i\rangle\langle i| $ signifies the projection onto the $ i$-th basis state.

We define the coherence measure $\mathcal{C}$ for a state $\sigma $ according to the following criteria:

1. $\mathcal{C}(\sigma)$ equals zero if and only if $\sigma$ belongs to the set of density matrices $\mathcal{I}$.

2. The coherence measure $\mathcal{C}(\sigma)$  exhibits monotonic behaviour under incoherent selective measurements: $\mathcal{C}(\sigma)$  is greater than or equal to $ \sum_{i} p_i C(\sigma_i) $, where$\sigma_i = \mathcal{E}_i \sigma \mathcal{E}_i^{\dagger}$ and $ p_i=Tr (\mathcal{E}_i \sigma \mathcal{E}_i^{\dagger} )$. Here, $ \mathcal{E}_i$ denotes a set of operators satisfying $\sum_{i}  \mathcal{E}^\dagger_i\mathcal{E}_i = 1 $ and $ \mathcal{E}_i \mathcal{I} \mathcal{E}_i^\dagger \subseteq \mathcal{I}$.

3. The coherence measure $\mathcal{C}$ exhibits convexity, that is  $\mathcal{C}(\sum_{i} p_i \sigma_i) \le \sum_{i} p_i \mathcal{C}(\sigma_i) $ for arbitrary states  $\{\sigma_i, p_i \} $, $p_i$ is the probability of occurrence of $\sigma_i$.

	Based on these axioms,  $l_1$-norm  has been shown to be a valid measure of coherence and is defined for a given quantum  state $\sigma$ as follows: 
	
	\begin{eqnarray}
		\mathcal{C}_{l_1}= \underset{\rho_{\mathcal{I}} ~\epsilon~ \mathcal{I}}{min}|| \sigma-\rho_{\mathcal{I}}||_{l_1}. 
	\end{eqnarray}
Next, optimization is carried out over all  $\rho_{\mathcal{I}}$ within the set  $\mathcal{I}$ and the above expression for $l_1$-norm  is reduced to the following form in the standard basis $\{|m\rangle\}$:
	\begin{eqnarray}
		\label{coh}
		\mathcal{C}_{l_1} = \sum_{m\ne n}|\langle m |\sigma |n\rangle|.
	\end{eqnarray}
This implies 	that the coherence given by  $l_1-$norm  $\mathcal{C}_{l_1}$ is simply related to the off-diagonal elements of the density matrix and is given by the  sum of  magnitudes of the off-diagonal elements of a given density matrix. Next, from the information point of view, any deviation from the monotonic behaviour of coherence is a signature for the non-Markovian dynamics. Let $\Delta \sigma (t)= \frac{d\mathcal{C}_{l_1}}{dt}$ be the time derivative of coherence, the Markovian dynamics is given by monotonicity of the $\mathcal{C}_{l_1}$ i.e $\Delta \sigma (t) \le 0 $. Thus $\Delta \sigma (t)>0$ for any time interval signifies non-Markovianity.  Based on this notion of back-flow of information, the amount of non-Markovianity is defined as \cite{coh2,col1}
	\begin{eqnarray}
		\label{Ncoh}
		\mathcal{N}_{\mathcal{C}_{l_1}}= \underset{\rho(0)}{ max}  \underset{\Delta \sigma (t) > 0}{\int}\!\!\!\!\!\!\! dt~ \Delta \sigma (t) 
	\end{eqnarray}
	where the maximization is done over initial coherent states and integration is carried out only for $\Delta \sigma (t) >0$ to access non-Markovian dynamics.

	\section{Spin Boson Model}
	In this section, we introduce spin boson model. We consider a qubit modeled by a two level system coupled strongly to a bosonic mode of energy $\omega$ ($\hbar=1$) \cite{sm3}:
	\begin{eqnarray}
		\label{model}
	\mathcal{	H}=J \sigma^x + \omega a^{\dagger} a + g \omega \sigma^z ( a +  a^{\dagger}),
	\end{eqnarray}  
	where $J$ is the tunneling energy and  $g $ is the qubit-bath coupling. $ \sigma^x$, $\sigma^y$, $ \sigma^z$ are the Pauli spin matrices. $a$ and $a^{\dagger}$ are annihilation and creation operators for the bosonic mode.
	This fundamental model plays a critical role across various scientific disciplines, including condensed matter physics, quantum optics, and quantum information. In the realm of quantum optics, it finds application in describing the interaction between an atom and a specific electromagnetic field mode  \cite{JC}. Condensed matter physics leverages this model to understand electron-phonon coupling in diverse systems, encompassing the Holstein model \cite{HM}, semiconductors \cite{semi} and superconductors \cite{sup}. Its applicability extends further to mechanical oscillator systems  \cite{mech}. Notably, the model also serves as a theoretical tool for analyzing macroscopic oscillations observed in superconducting systems \cite{sm}.

	Next, in order to make perturbative calculations, we transform the Hamiltonian in above equation \ref{model} to polaron frame using Lang-Firsov transformation \cite{amy,ms}. Defining, 
	$	S=-g\sigma^{z}(a-a^{\dagger})$, so that the transformed Hamiltonian is given by
	\begin{eqnarray}
	\mathcal{	H}_p =e^{-S}\mathcal{H} e^{S}= \mathcal{H}_S + \mathcal{H}_B +\mathcal{H}_I,
	\end{eqnarray}
	where in the transformed frame the qubit gets dressed by phonons-known as polaron and  the transformed Hamiltonian for the system is $H_S= \tilde{J} \sigma^x$ with modified tunneling rate $\tilde{J}=Je^{-2g^2}$.  The bath Hamiltonian in the polaron frame   $H_B=\omega a^{\dagger} a$, while the interaction Hamiltonian is
	
	\begin{eqnarray}
	\mathcal{	H}_I &=& \tilde{J}[ \sigma^{-} (e^{-2g\omega b^{\dagger}} e^{2g\omega b}-1 ) + \sigma^+ (e^{2g\omega b^{\dagger}} e^{-2g\omega b}-1 )],\nonumber\\
		&=& \tilde{J}[ \sigma^{-} \mathcal{F} ^{\dagger}+ \sigma^+ \mathcal{F}],
	\end{eqnarray}
	with $\sigma^{\pm}=\frac{\sigma^x \pm i \sigma^y}{2}$ are system ladder operators and  $ \mathcal{F}^\dagger= (e^{-2g\omega b^{\dagger}} e^{2g\omega b}-1 )$ represents modified bath operators coupled to the system in polaron frame. The coupling in polaron frame gets reduced to the value $\frac{Je^{-2g^2}}{\omega}$ and is small in comparison to the bare coupling $g$, thus perturbative calculation can be done.
	
	{ Now we sum up the different time scales that arise in our model. From equation \ref{model}, we have two time scales defined by adiabaticity parameter $\frac{J}{\omega}$, and the interaction scale $\frac{g\omega}{\omega}=g$. If $\frac{J}{\omega}<<1$, we say the dynamics is anti-adiabatic. However, in the polaron frame (after Lang-Firsov transformation), these scales change to $\frac{Je^{-2g^2}}{\omega}$ and $Je^{-2g^2}$ respectively. We note that the bath correlation time scale is given by $\tau_B \sim \frac{1}{\omega}$, while system relaxation time scale is given by $\tau_R \sim \frac{1}{Je^{-2g^2}}$. Thus, if $\tau_B >>\tau_R$ which implies $\frac{Je^{-2g^2}}{\omega}<<1$, the dynamics is  Markovian, while the anti-adiabatic condition in polaron frame is the same condition $\frac{Je^{-2g^2}}{\omega}<<1$, (see the discussion at the end of next section). }
	
	\section{Non-Markovianity in the  dynamical evolution}

In order to investigate the non-Markovian effects in the dynamics of aforementioned   model, we employ the time-convolutionless (TCL) master equation, which effectively captures non-Markovian dynamics within the perturbative regime. Let $\rho^T(0)$ be the initial density matrix for the total system and bath, therefore, in the Born approximation that assumes  weak system-bath coupling,  we can write it in the form $\rho^T(0)=\rho_S(0) \otimes \rho_B$, where the qubit density matrix is given by  $\rho_S(0) $,  and $ \rho_B = \frac{e^{-\beta H_B}}{Z_B} $ denotes the density matrix associated with the bath, where $ Z_B $ represents the partition function of the bath. In the polaron frame, the qubit-bath coupling is weak, therefore, we 	 directly apply the TCL master equation as follows \cite{BP}:
	
	\begin{eqnarray}
	\label{mas}
	\dot{\rho}_S^I(t) =
	-i {\rm Tr_B}[\mathcal{H}^I(t), \rho^T(0)] - \int_0^t d\tau {\rm Tr_B}[\mathcal{H}^I(t),[\mathcal{H}^I(\tau), \rho_S^I(t) \otimes \rho_B]] 
\end{eqnarray}
	Where $\dot{\rho}_S^I(t)$ means time derivative of  $\rho_S(t)$;  $ \rho_S^I(t)$ is the qubit density matrix  defined in the interaction picture, ${\rm Tr}_B$ means trace over bath, and   $\mathcal{H}^I(t)$ is the interaction Hamiltonian in the interaction picture.  The operators in interaction picture are defined as: $\hat{O}^I(t)= e^{i\mathcal{H}_0 t}\hat{O} e^{-i\mathcal{H}_0t}$,  $\mathcal{H}_0$ is the free part of the total Hamiltonian.  In order to evaluate each  term in the this equation, we first simply $\mathcal{H}_I(t)$. Let $\{|E_n\rangle\}$ be the  energy  eigen basis for the system and $\{|n_{ph}\rangle\}$ be the energy  eigen basis for phonons with energies $\omega_n$, therefore we write
	\begin{eqnarray}
		\label{LFH}
		\mathcal{H}^I(t) &=& e^{-i\mathcal{H}_0 t} \mathcal{H}_I e^{i \mathcal{	H}_0 t}\nonumber \\
		& =&\sum_i  |E_i\rangle \langle E_i| \sum_m |m_{ph}\rangle \langle m_{ph}| e^{-i(\mathcal{H}_S+\mathcal{H}_B)t }\mathcal{H}_I e^{i(\mathcal{H}_S+\mathcal{H}_B)t } \sum_j  |E_j\rangle \langle E_j| \sum_n |n_{ph}\rangle \langle n_{ph}|  \nonumber \\
		&=& \sum_{i,j} \sum_{m,n} e^{-i[(E_i-E_j) + (\omega_m-\omega_n)]t} |E_i \rangle \langle E_j| |m_{ph}\rangle \langle n_{ph}| \langle E_i| \langle m_{ph} |\mathcal{ H}_I |E_j\rangle |n_{ph}\rangle.
	\end{eqnarray}
	We see that $E_i-E_j \propto \tilde{J}=Je^{-2g^2}$ and $\omega_m-\omega_n \propto \omega$. If we make an assumption that $\frac{Je^{-2g^2}}{\omega} <<1$ ,
	then $(E_i-E_j+ \omega_m-\omega_n) \sim \omega [\frac{Je^{-2g^2}}{\omega}+1]\sim \omega  $ \cite{amy,ms}. Therefore, we can ignore the $E_i-E_j$ term in  exponential of the above equation. This  is also called  as an  anti-adiabatic approximation. This implies that there is no time evolution of system operators and therefore we write
	\begin{eqnarray}
		\label{ad}
	\mathcal{	H}^I(t)&=&  \sum_{m,n} e^{-i (\omega_m-\omega_n)t}  |m_{ph}\rangle \langle n_{ph}| \langle m_{ph} | \mathcal{H}_I  |n_{ph}\rangle = \tilde{J}[ \sigma^{-}\mathcal{F}^{\dagger}(t)+ \sigma^+ \mathcal{F}(t)],
	\end{eqnarray}
	where $\mathcal{F}^\dagger(t)= e^{i\mathcal{H}_Bt}\mathcal{F}e^{-i\mathcal{H}_Bt}= e^{-2g\omega b^{\dagger}e^{i\omega t}}e^{2g\omega be^{-i\omega t}}-1 $ is the time evolved modified bath operator. 
	This approximation can be understood in the following way. The relaxation time scale for the qubit is $\tau_R \sim \frac{1}{\tilde{J}}$ while the time scale over the bath correlations decay is $\tau_B \sim \frac{1}{\omega}$. The Markovian approximation implies $\tau_R>>\tau_B$, which means $\frac{Je^{-2g^2}}{\omega}<<1$. Thus anti-adiabatic approximation intrinsically implies Markovian dynamics in the system. Next, we restrict our calculation to zero temperature  case so that
	\begin{eqnarray}
		{\rm Tr_B}[\mathcal{H}^I(t) \rho^T(0)]&=&\sum_{n} \langle n_{ph}| \mathcal{H}^I(t) \rho_S(0)  \rho_B |n_{ph}\rangle\nonumber \\
		&=& \frac{1}{Z_B}\sum_{n} \langle n_{ph}| \mathcal{H}^I(t) \rho_S(0)  |n_{ph}\rangle e^{-\beta \omega_n}  \\
		& \overset{\beta \rightarrow \infty}{=}&  \langle 0_{ph}| \mathcal{H}^I(t) \rho_S(0)  |0_{ph}\rangle=  \langle 0_{ph}|\tilde{J}[ \sigma^{-}\mathcal{F}^{\dagger}(t)+ \sigma^+ \mathcal{F}(t)]|0_{ph}\rangle \rho_S(0)=0
		\label{first}
	\end{eqnarray}
	where we have used for $T\rightarrow 0$, the bath is in the vacuum state $|0_{ph}\rangle$. Thus at zero temperature, the first term in master equation \ref{mas} vanishes. 
	\subsection{Naive Markovian dynamics}
	Using the above approximation $\frac{Je^{-2g^2}}{\omega}<<1$ and the result in equation \ref{first}, the time limit in the above master equation \ref{mas} can be taken to $\infty$ with $\tau$ replaced by $t-\tau$ \cite{BP}, so that we can write 
	
	\begin{eqnarray}
		\label{masi}
		\dot{\rho}_S^I(t)&=& -\int_0^{\infty} d\tau {\rm Tr_B}[\mathcal{H}^I(t),[\mathcal{H}^I(t-\tau), \rho^I_S(t) \otimes \rho_B]].
	\end{eqnarray}
This change in the limit amounts to ignore the leading order corrections to this master equation. Here, we show  according to  this limit, the anti-adiabatic approximation yields no decoherence in the system . However, in the next subsection, we show some  non-Markovian behaviour exists for certain coupling range.  Now, evaluate each term in the double commutator above. It contains four terms, the first one can be written as below:

	\begin{eqnarray}
		{\rm Tr_B}[\mathcal{H}^I(t)\mathcal{H}^I(t-\tau) \rho^I_S(t) \rho_B]
		&=& \frac{1}{Z_B}\sum_n e^{-\beta \omega_{n}} \langle n_{ph}| \mathcal{H}^I(t)\mathcal{H}^I(t-\tau)|n_{ph}\rangle \rho^I_S(t)\nonumber\\
		&\overset{\beta \rightarrow \infty}{=}& \langle 0_{ph}|\mathcal{H}^I(t)\mathcal{H}^I(t-\tau)|0_{ph}\rangle \rho^I_S(t) \nonumber \\
		&=& \sum_{n}  \langle 0_{ph}| H^I(t)|n_{ph}\rangle \langle n_{ph}| \mathcal{H}^I(t-\tau)|0_{ph}\rangle\rho^I_S(t),\\
		&=& \sum_{n}  \langle 0_{ph}| e^{i\mathcal{H}_B t} \mathcal{H}_I e^{-i\mathcal{H}_B t}|n_{ph}\rangle \langle n_{ph}| e^{i\mathcal{H}_B (t-\tau)}\mathcal{H}_Ie^{-i\mathcal{H}_B (t-\tau)}|0_{ph}\rangle \rho^I_S(t), \\
		&=& \sum_{n}  e^{i\omega_n t} |\langle 0_{ph}|\mathcal{H}_I|n_{ph}\rangle|^2\rho^I_S(t),
	\end{eqnarray}
	where in the third equality, we have used equations \ref{LFH} and \ref{ad}. Similarly, we can evaluate other terms. Therefore, the master equation simplifies to
	
	\begin{eqnarray}
	\dot{\rho}_S^I(t)
		&=& -  \sum_n  
		\left[ \underset{\zeta \rightarrow 0^+}{{\rm lim}} \Big[ \int_0^{\infty} d\tau ~ e^{- i(\omega_n -i \zeta)\tau} |\langle 0_{ph} |\mathcal{H}_I |n_{ph} 
		\rangle|^2 {\rho}^I_S(t)  + 
		\int_{0}^{\infty} d\tau ~ e^{i(\omega_n+i\zeta)\tau}
		~{\rho}^I_S(t) ~|\langle 0_{ph} | \mathcal{H}_I |n_{ph} \rangle|^2 \Big] \right .\nonumber \\
		&&~~~~~~~~~~~~~~ - \left. \int_{-\infty}^{\infty} d\tau ~ e^{i\omega_n \tau} ~ \langle n_{ph}| \mathcal{H}_I 
		|0_{ph}\rangle  	~{\rho}^I_S(t)~ \langle 0_{ph}| \mathcal{H}_I|n_{ph}\rangle 
		\right]
		\label{21}
	\end{eqnarray}
	Using the identity  $\int_{-\infty}^{\infty} dx e^{i z x} = 2\pi \delta(z)$ and $\delta(z-a)f(z)= \delta(z-a)f(a)$, the term 
	\begin{eqnarray}
		\int_{-\infty}^{\infty} d\tau ~ e^{i\omega_n \tau} \langle n_{ph}| \mathcal{H}_I
		|0_{ph}\rangle {\rho}^I_S(t)\langle 0_{ph}|\mathcal{H}_I|n_{ph}\rangle &=& 2 \pi \delta(\omega_n) \langle n_{ph}|\mathcal{H}_I
		|0_{ph}\rangle  	~{\rho}^I_S(t)~ \langle 0_{ph}|\mathcal{H}_I|n_{ph}\rangle \\
		&=&  \frac{2 \pi}{\omega}\delta(n) \langle 0_{ph}| \mathcal{H}_I
		|0_{ph}\rangle 	~{\rho}_S(t)~ \langle 0_{ph}| \mathcal{H}_I|0_{ph}\rangle=0,
	\end{eqnarray}
where we have used $\omega_{n}=n\omega$ and $\delta(\omega_{n})=\frac{1}{\omega}\delta(n)$.	The master equation \ref{21} reduces to 
	\begin{eqnarray}
		\label{22}
		\frac{d {\rho}^I_S(t)}{dt} 
		&=& i \sum_n  \frac{1}{\omega_n} 
		\left[  |\langle 0_{ph} |\mathcal{H}_I |n_{ph} 		\rangle|^2 {\rho}^I_S(t)  -
		~{\rho}^I_S(t) |\langle 0_{ph} |\mathcal{H}_I |n_{ph} \rangle|^2 \right].
	\end{eqnarray}
	Next, we have
	
	\begin{eqnarray}
		\sum_n  \frac{1}{\omega_n} 	|\langle 0_{ph} |\mathcal{H}_I |n_{ph} 	\rangle|^2 &=&  \sum_n  \frac{1}{\omega_n} \langle 0_{ph} | \mathcal{H}_I|n_{ph} 	\rangle\langle n_{ph} | \mathcal{H}_I |0_{ph}	\rangle\\
		&=&  \sum_n  \frac{1}{\omega_n} | \langle 0_{ph} | \mathcal{F}|n_{ph} 	\rangle|^2 [\sigma^+ \sigma^- + \sigma^-\sigma^+  ] = \sum_n  \frac{1}{\omega_n} | \langle 0_{ph} | \mathcal{F}|n_{ph} 	\rangle|^2,
	\end{eqnarray}
	with $\sigma^+ \sigma^- + \sigma^-\sigma^+ =1$. Using this result, the time derivative of density matrix \ref{22} reduces to 
	\begin{eqnarray}
		\frac{d {\rho}^I_S(t)}{dt} &=& \sum_n  \frac{1}{\omega_n}	\left[  | \langle 0_{ph} | \mathcal{F}|n_{ph} 	\rangle|^2  {\rho}^I_S(t)  -
		~{\rho}^I_S(t) | \langle 0_{ph} | \mathcal{F}|n_{ph} 	\rangle|^2 \right] = 0 \nonumber \\
		&& \implies\rho^I_S(t)=\rho^I_S(0).
	\end{eqnarray}
	This means the system is decoherence free! This is inconsistent due to the fact the qubits still undergo decoherence in Markovian approximations. This inconsistency occurs due to changing limits of the time integration in the master equation above. Keeping the finite time limits unveils the non-Markovian behaviour discussed next.

	\subsection{Finite time evolution}
	
\begin{figure}[t]
		\includegraphics[width=4.5cm,height=4.5cm]{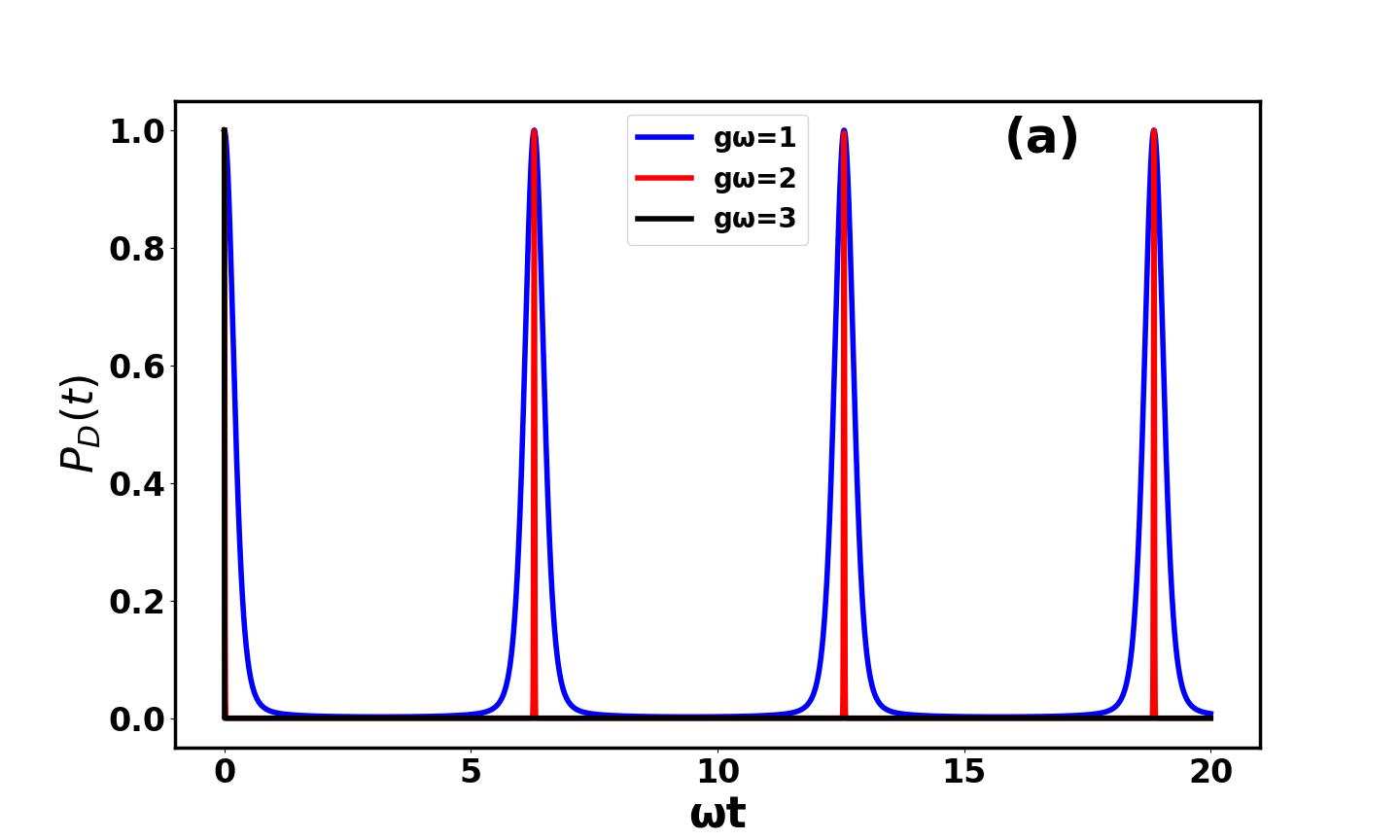}\hspace{0.5cm}
		\includegraphics[width=4.5cm,height=4cm]{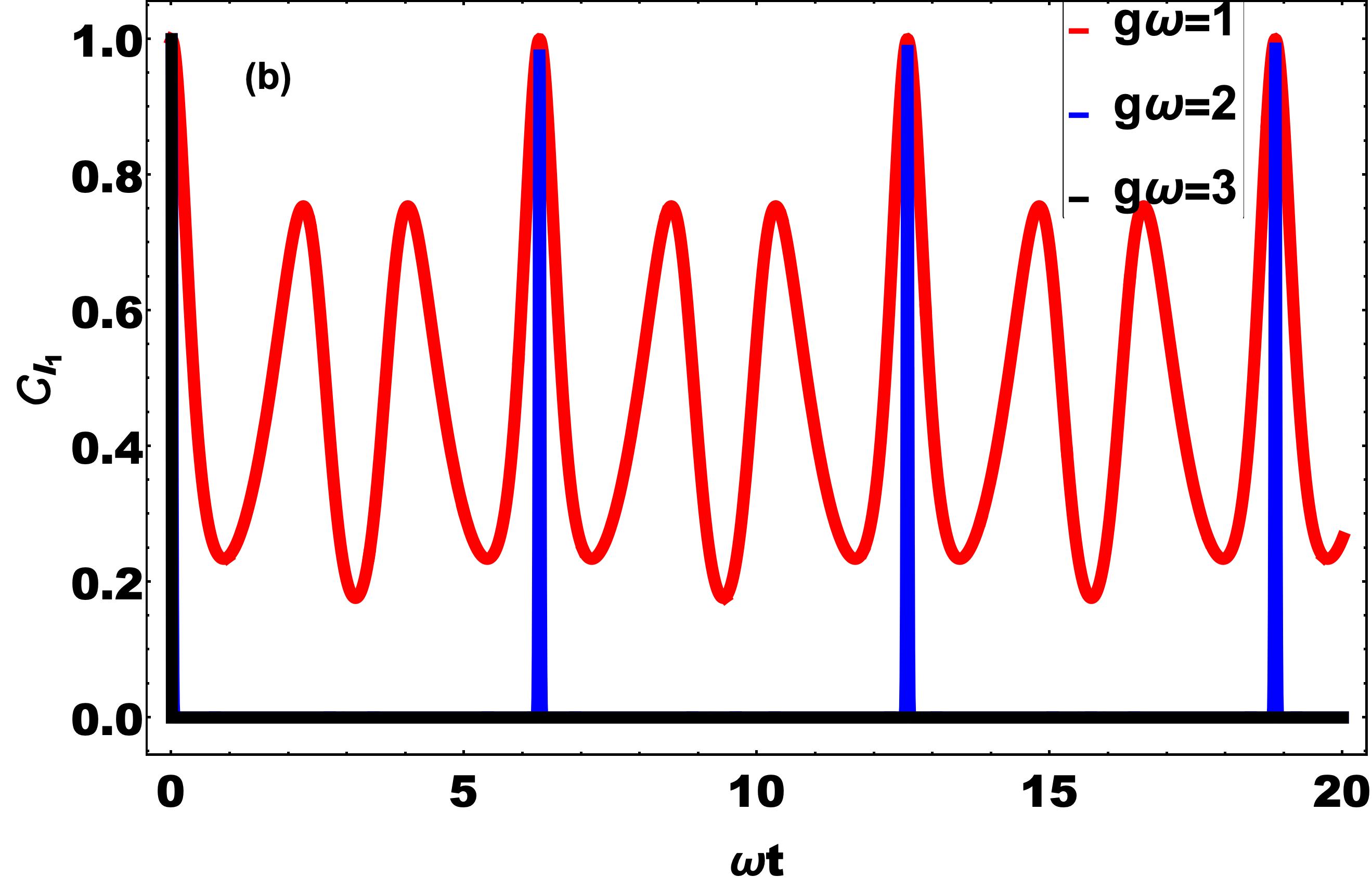}\hspace{0.5cm}
		\includegraphics[width=4.5cm,height=4.1cm]{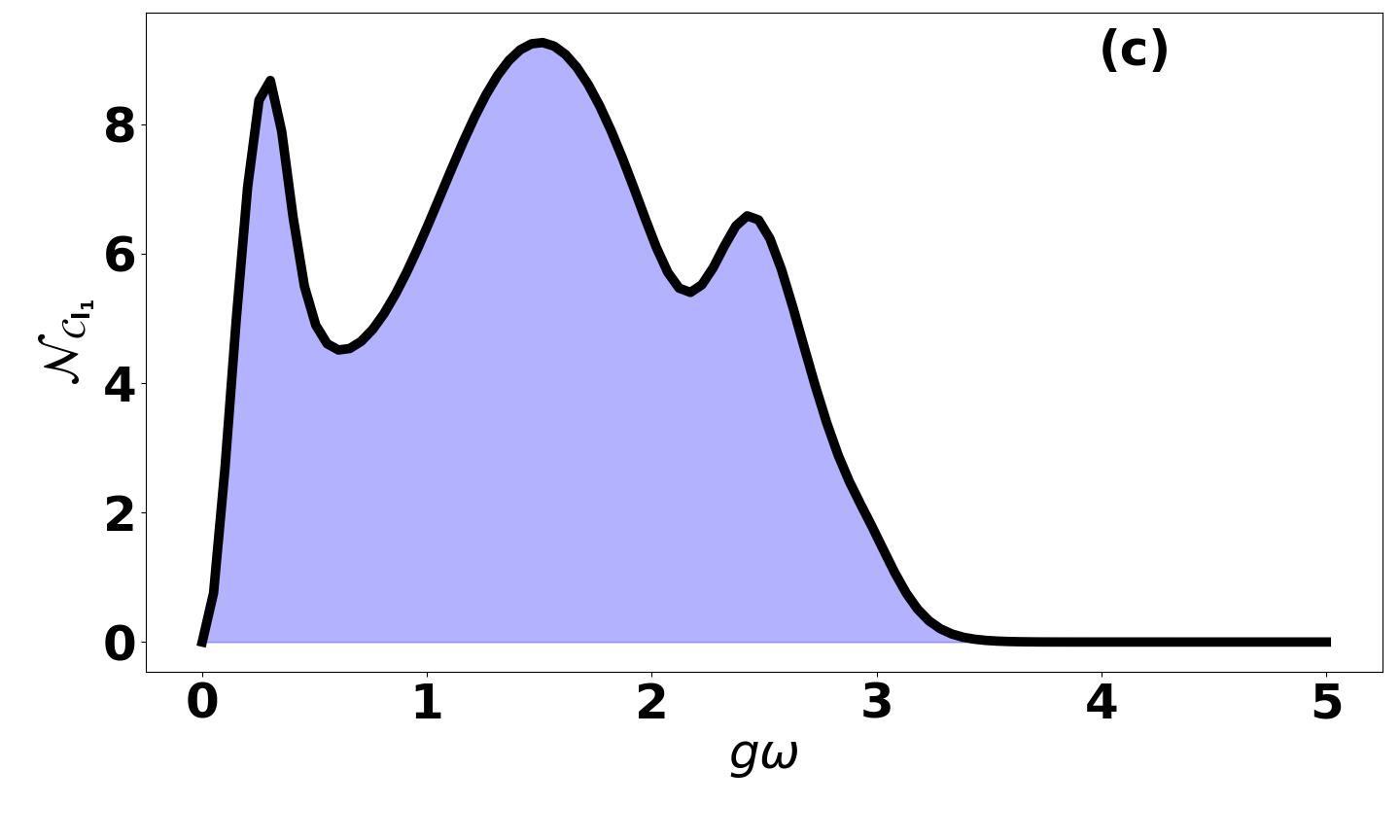}
		\caption{In this  graph (a) represents the time variation of population difference $P_D(t)$ with respect to $\omega t$ as given in equation \ref{PDT}. We see that there is a revival of $P_D(t)$ at certain instants of time reflecting non-Markovian behaviour. In (b) we plot coherence $C_{l_1}(t)$ vs $\omega t$ at zero temperature for different values of $g\omega$, (c) non-Markovianity $\mathcal{N}_{C_{l_1}}$ vs coupling $g\omega$. We see that there exist finite non-Markovianity (shaded region) for  certain coupling range while it vanishes for large couplings beyond value of 3 }
		\label{PD}
	\end{figure}

	Now, instead of changing integration limit in equation \ref{mas} to infinity, we keep it finite and still assume  $\frac{Je^{-2g^2}}{\omega}<<1$.
	Defining $\alpha_1(t-\tau) = \tilde{J}^2\langle \mathcal{F}^{\dagger}(t)\mathcal{F}^\dagger(\tau)\rangle_B $ and $\alpha_2(t-\tau)= \tilde{J}^2 \langle \mathcal{F}^{\dagger}(t)\mathcal{F}(\tau)\rangle_B $, we write the above master equation at finite temperature as follows (see appendix A for details):
	\begin{eqnarray}
		\frac{d \rho^I_S(t)}{dt}&=&\int_0^t d\tau \Bigg[\alpha_1(t-\tau)[     \sigma^+ \rho^I_S(t) \sigma^+ + \sigma^- \rho^I_S(t) \sigma^-]  \nonumber \\
		&&~~~~~~~~~	+\alpha_2(t-\tau)[ \sigma^+\rho^I_S \sigma^- - \sigma^- \sigma^+ \rho^I_S(t) - \sigma^+ \sigma^- \rho^I_S(t)+ \sigma^- \rho^I_S \sigma^+      ]    + {\rm h.c.}                    \Bigg],
	\end{eqnarray}
	where $h.c.$ means Hermitian conjugate.  Since, we assume in our calculations at zero temperature  i.e.  $\beta \rightarrow \infty$ limit, so that we can write this equation in the Schrodinger picture as:  
	\begin{eqnarray}
		\label{mas2}
		\frac{d\rho_S}{dt}&=&-i[\mathcal{H}_{S},\rho_S]+\beta_+(t)[\sigma^{+}\rho_S\sigma^{+}+\sigma^{-}\rho_S \sigma^{-}]\nonumber\\
		&&+~\beta_-(t)\{[2\sigma^{+}\rho_S\sigma^{-}-\{\sigma^{-}\sigma^{+}, \rho_S\}]+[2\sigma^{-}\rho_S\sigma^{+}-\{\sigma^{+}\sigma^{-},\rho_S\}]\},
	\end{eqnarray}
	where $\beta_{\pm}(t)= 2\tilde{J}^2\sum_{n=1}^{\infty}\frac{(\pm 4g^2\omega^2)^n}{n!}\frac{\sin n\omega t}{n\omega}$. We assume initial density matrix of the qubit to be of the form 
		$\rho_S(0)= \begin{pmatrix}
			\rho_{00} & \rho_{01} \\
			\rho_{01} & \rho_{11}
		\end{pmatrix} $ in  standard basis $\{|0\rangle, |1\rangle\}$. Then according to the above master equation \ref{mas2}, the elements of this density matrix evolve as follows:
	\begin{eqnarray}
		\frac{d}{dt} \begin{pmatrix}
			\rho_{00}(t) \\
			\rho_{01}(t)\\
			\rho_{10}(t)\\
			\rho_{11}(t)
		\end{pmatrix}
 		= \begin{pmatrix}
 			-2\beta_+(t) & 0& 0& 2\beta_+(t)\\
 		0& -2\beta_-(t)	& 2\beta_+(t) & 0\\
 		0& 2\beta_+(t) & -2\beta_-(t) & 0	\\
 			2\beta_+(t) & 0& 0& -2\beta_+(t)
  		\end{pmatrix}
  		\begin{pmatrix}
  			\rho_{00}(t) \\
  		\rho_{01}(t)\\
  		\rho_{10}(t)\\
  		\rho_{11}(t)
  		\end{pmatrix}
 		\label{offdi}
	\end{eqnarray}
	These equations can be solved exactly. We see that ${\rm Tr_S} \rho_S(t)=\rho_{00}(t)+\rho_{11}(t)=1$, i.e. trace is preserved. The  difference in populations $P_D$ for $|0\rangle$ and $|1\rangle$ therefore evolve as
	\begin{eqnarray}
		\label{PDT}
		P_D(t)=\frac{\rho_{00}(t)-\rho_{11}(t)}{\rho_{00}(0)-\rho_{11}(0)}= e^{-\Gamma(t)},
	\end{eqnarray}
	where $\Gamma(t)= -4\tilde{J}^2 \sum_{l=1}^{\infty} \frac{(4g^2\omega^2)^l}{l!} \frac{\cos l\omega t-1}{l^2 \omega^2}$. In figure \ref{PD} (a), we plot the $P_D(t)$ for different values of coupling $g\omega$. We observe that there are strong oscillations of $P_D(t)$ indicating non-Markovian behaviour.  Furthermore, we see that if the system is localized in some state initially (say $|0\rangle $), it gets delocalized abruptly into a superposition state of $\ket{0}$ and $\ket{1}$ with equal probabilities. At  $t=\frac{2\pi m}{n \omega}=\frac{2\pi}{\omega} \mathcal{I}^+$ ($m$, $n$ are integers and $\mathcal{I}^+$ is a positive integer), we see the system goes back into the original state. This behaviour is related to coherent-incoherent transition of spin-boson model \cite{Leg}.

	\subsection{Coherence}
	
	In this section, we assess the non-Markovianity in the dynamical evolution of the qubit through coherence defined in equations \ref{coh} $\&$ \ref{Ncoh}. We write
	coherence in terms of $l_1$-norm at some time $t$ as 
	\begin{eqnarray}
		\label{coh1}
		\mathcal{C}_{l_1}= |\rho_{01}(t)| + |\rho_{10}(t)|.
	\end{eqnarray}
From the equations \ref{offdi}, the off-diagonal terms evolve according to
\begin{eqnarray}
	\label{offd}
	\rho_{01}(t)= \rho_{01}(0)[\frac{e^{-\Gamma_{even}(t)} + e^{-\Gamma_{odd}(t)}}{2}] + \rho_{10}(0)[\frac{[e^{-\Gamma_{even}(t)} - e^{-\Gamma_{odd}(t)}]}{2}],
\end{eqnarray}
where $\Gamma_{odd(even)}=-4 \tilde{J}^2 \underset{n=odd(even)}{\sum} \frac{(4g^2\omega^2)^n}{n!} \frac{\cos n\omega t-1}{n^2 \omega^2}$. 
	Assuming, $\rho_{01}(0)=|\rho_{01}(0)|e^{i\phi}$, where $\phi$ is the relative phase. Therefore, we can simplify above equation \ref{coh1} using the result in equation \ref{offd} as
	\begin{eqnarray}
		\label{coh1}
		\mathcal{C}_{l_1}(t)=\mathcal{C}_{l_1}(0)\sqrt{\cos \frac{\phi}{2}e^{-2\Gamma_{even}} + \sin \frac{\phi}{2}e^{-2\Gamma_{odd}} },
	\end{eqnarray}
	where $ \mathcal{C}_{l_1}(0) =2|\rho_{01}(0)|$ is the initial coherence in the system. The derivative of $\mathcal{C}_{l_1}(t)$ is given by 
	\begin{eqnarray}
		\Delta(t) = \frac{d\mathcal{C}_{l_1}(t)}{dt}= -\frac{\mathcal{C}_{l_1}^2(0)}{\mathcal{C}_{l_1}(t)} \Bigg[\cos \frac{\phi}{2} \frac{d\Gamma_{even}}{dt}+ \sin \frac{\phi}{2} \frac{d\Gamma_{odd}}{dt}\Bigg].
	\end{eqnarray}
	Therefore, non-Markovianity is given by equation \ref{Ncoh} 
	\begin{eqnarray}
		\label{ncoh1}
		\mathcal{N}_{\mathcal{C}_{l_1}}= \underset{\rho_S(0)}{{\rm max}}\int_{\Delta(t) >0} dt \Delta(t).  
	\end{eqnarray}
	Now, we plot $	\mathcal{C}_{l_1}(t)$  and $	\mathcal{N}_{\mathcal{C}_{l_1}} $ in figure \ref{PD} (b) and figure \ref{PD}(c). Without loss of generality, we take $\phi=0$, so that initial state are the eigen states of $\sigma^x$ i.e. $|\pm\rangle= \frac{1}{2}[\ket{0}\pm \ket{1}]$. In such a case, $	\mathcal{C}_{l_1}(t)=	\mathcal{C}_{l_1}(0)e^{-\Gamma_{even}(t)}$. In figure \ref{PD}(b), we have  $\mathcal{C}_{l_1}(t)$ versus $t$ for different values of $g\omega$. We observe that there are strong oscillations of coherence indicating non-Markovian behaviour. For $g\omega=1$, we see that the coherence does not go to zero while as for $g\omega \ge 2$, it  goes to zero within certain time intervals and revives at particular instants of time. This abrupt vanishing of coherence and then revival is in the same line as entanglement sudden death and revival . Furthermore, these sharp peaks occur at the instants of time  when $\Gamma_{even}(t)=0$ which happen if $t=\frac{2\pi}{\omega} \frac{m}{n} =\frac{2\pi}{\omega} \mathcal{I}^+ $, for exactly the ratio $\frac{m}{n}$ to equal to $\mathcal{I}^+ $ a positive integer ($m,n$ are integers). The sharp peaks are attributed to the single mode nature of the bath. However, it can be shown that the same behaviour occur for multimode bath also  but with a broad peaks. Thus these instants of time mark the non-Markovian effects  in the system dynamics.     Next, in order to measure the degree of non-Markovianity, we plot $\mathcal{N}_{\mathcal{C}_{l_1}}$ with respect to $g\omega$ in \ref{PD}(c) . We see that there is finite amount of non-Markovianity in the dynamics for certain coupling range, thus invalidating the notion of Markovian dynamics in the polaron frame for large polarons. However,  $\mathcal{N}_{\mathcal{C}_{l_1}}$ vanishes for $g\omega\ge 3$, thus implying  system dynamics is  Markovian. Therefore, in the very large coupling limit, the system dynamics can be taken as  Markovian.

{
	
 Next, in order to understand the physical processes that govern the above dynamics,  we  examine the various  approximations involved above. First, we try to figure out the order of magnitude of the various terms involved in going from equation \ref{mas} to equation \ref{masi}. It suffices to consider only one term in equation \ref{mas} and under Markovian approximation we write
 \begin{eqnarray} 
 	\int_0^{t}d\tau \sigma^+(t) \rho_S(t)  \sigma^-(\tau) \mathcal{\alpha}(t, \tau)	&\rightarrow&  \int_0^{\infty}d\tau  \sigma^+ \rho_S(t)  \sigma^- \mathcal{\alpha}(t-\tau) +   {\rm corrections}.
 \end{eqnarray}
Here $\mathcal{\alpha}(t)$ is the bath correlation function and  we assume its form to be $e^{-\omega t}$, so that bath correlation time scale is $\tau_B \sim \frac{1}{\omega}$. The order of magnitude of the first term on r.h.s i.e. the term used as Markovian approximation is $\tau_B \sim \frac{1}{\omega}$: 
$ ||  \int_0^{\infty}d\tau  \sigma^+ \rho_S(t)  \sigma^- \mathcal{\alpha}(t-\tau) || \le  \int_0^{\infty}d\tau   |\mathcal{\alpha}(t-\tau)|\sim \frac{1}{\omega} =O(\tau_B)$. $||...||$ defines trace norm and its identities are used \cite{fun1,fun2}. Next, we look at the order of magnitude of the leading order of the corrections, that can be simply looking at the upper bound of  the first term. We write, using triangle inequality
\begin{eqnarray}
||	\int_0^{\infty}d\tau  \sigma^+ [\rho_S(t) -\rho_S(t-\tau) + \rho_S(t-\tau) ] \sigma^- \mathcal{\alpha}(t-\tau)||&  \le &
||	\int_0^{\infty}d\tau  \sigma^+ [\rho_S(t) -\rho_S(t-\tau) ] \sigma^- \mathcal{\alpha}(t-\tau)|| \nonumber\\
&  \le & 	\int_0^{\infty}d\tau   || \tau \frac{[\rho_S(t) -\rho_S(t-\tau) ]}{\tau}||~  |\mathcal{\alpha}(t-\tau)| \nonumber \\
&=&\int_0^{\infty}d\tau    \tau ||\frac{ d\rho_S(t)}{d\tau}||~  |\mathcal{\alpha}(t-\tau)| \sim O(\tilde{J}^2 \tau^3) =O(\frac{\tilde{J}^2}{\omega^3}), 
\end{eqnarray}
where it can be shown that  $||\frac{ d\rho_S(t)}{d\tau}|| \sim O(\frac{\tilde{J}^2}{\omega} )$. Thus we see that the relative order of magnitudes of the leading correction term with the first one is $O((\frac{\tilde{J}}{\omega})^2)$. Therefore, in the approximation $\frac{Je^{-2g^2}}{\omega}<<1$, the leading order corrections can be ignored and we recover Markovian dynamics as in equation \ref{masi}. The finite time evolution given in equation \ref{mas} will thus resolve the dynamics at this order. Thus within the finite time evolution in the limit $\frac{Je^{-2g^2}}{\omega}<<1$  which although is a small effect, we have some non-Markovinity [figure 1(c)] that vanishes for larger bare couplings $g\omega$.

\begin{figure}[t]
	\includegraphics[width=5.5cm,height=5.5cm]{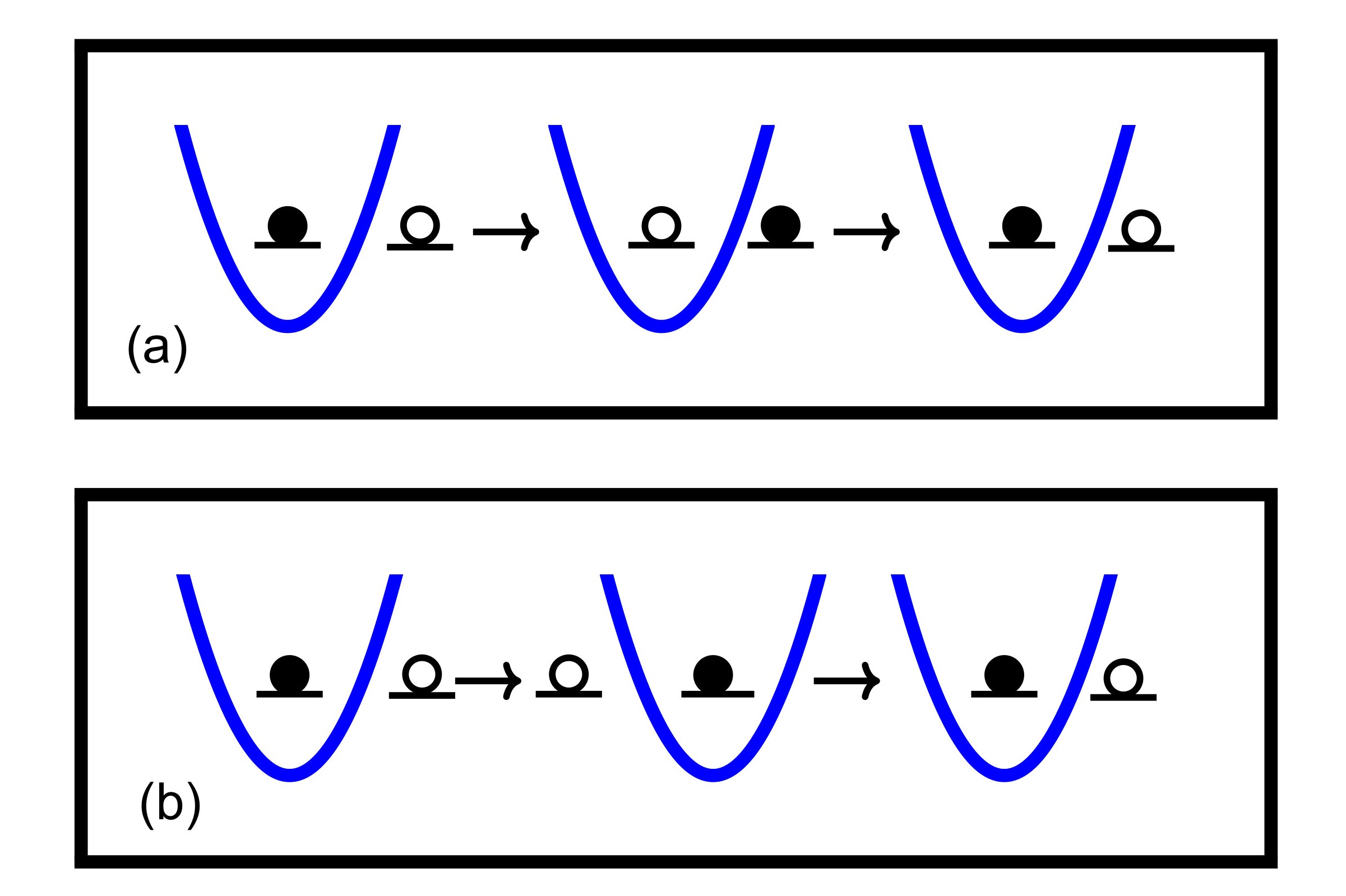}
		\caption{We consider only these two processes for the sake of discussion out of many different processes that occur at second order of perturbation. In each diagram, the left side represent initial state, middle one as the intermediate state while the right side is the final state. The particle starts in the initial state and returns back through some intermediate state.  A shaded circle means particle is present at that site and vice-versa. The parabolic wells represent the phonon modes (lattice distortion) with energy $-g^2 \omega$ ($+g^2\omega$)if particle is present (absent). }
	\label{Polaronfig}
\end{figure}

Now we try to correlate above discussion with the physical processes at second order of perturbation \cite{Pol1,Pol2}. We can consider our model as a double well potential ( or a two site problem) with one particle hopping between the lobes of double well potential (or two sites) while the phonons can be considered as some lattice distortion. In the transformed frame, we have a particle dressed with phonons-a polaron hopping with the effective rate $\frac{1}{\tilde{J}}$. Thus, we have two time scales for the particle hopping between the sites: (a) the time scale governing the full lattice distortion or the full relaxation of the polaron i.e. $\frac{1}{\tilde{J}}$ (b) time scale governing the negligible relaxation or distortion of the lattice: $\frac{1}{J}$. There are various processes occurring at second order perturbation, for the sake of discussion we keep only two of them [see figure \ref{Polaronfig}]. These processes can described by the hopping of the particle from one site to another with or without the relaxation or distortion of the lattice sites. Here, in this figure \ref{Polaronfig}, a shaded circle means particle is present at that site while empty site is represented by  a circle without a shade. The parabolic wells represent the phonon modes (lattice distortion) with energy $-g^2 \omega$ ($+g^2\omega$)if particle is present (absent). Next, in the processes shown in  the figure \ref{Polaronfig}(a),   the particle tunnels back to the original position through some intermediate states as allowed at second order of perturbation. In the intermediate state, the particle moves to second site leaving the first site with full lattice distortion and then finally going to original site without creating any new distortion. Thus during this process we conclude all these states have the same lattice distortion. This process occurs therefore at the rate $\tau_{R1}^{-1}= \frac{J}{g^2\omega}\times \tilde{J}= \frac{J^2 e^{-2g^2}}{g^2\omega}$. Now, in the figure \ref{Polaronfig}(b), we can have the process where the particle hops as a polaron from one site to other and then back. This occurs at the rate $\tau_{R2}^{-1}= \frac{\tilde{J}}{g^2\omega}\times\frac{\tilde{J}}{g^2\omega}=\frac{J^2e^{-4g^2}}{g^4\omega^2} $. Thus the relative rate of these process are 
$\frac{\tau_{R2}^{-1}}{\tau_{R2}^{-1}}= \frac{Je^{-2g^2}}{\omega}\frac{1}{Jg^2} \sim O(\frac{Je^{-2g^2}}{\omega}) $.  Thus in the limit $\frac{Je^{-2g^2}}{\omega}<<1$, there are several processes which are not resolved at this time scale by the master equation \ref{masi}. Therefore in the Markovian approximation, system has no influence on the bath as in first processes  while the vice-versa is true for  non-Markovian effects as in processes of type two. Therefore, a finite time evolution given by equation \ref{mas2} can resolve the contribution of these intermediate processes which shows the non-Markovian behaviour for a wide range of bare couplings $g\omega$.

 }
	
	\subsection{{Steady state limit}}
\begin{figure}[t]
	\includegraphics[width=5.8cm,height=5cm]{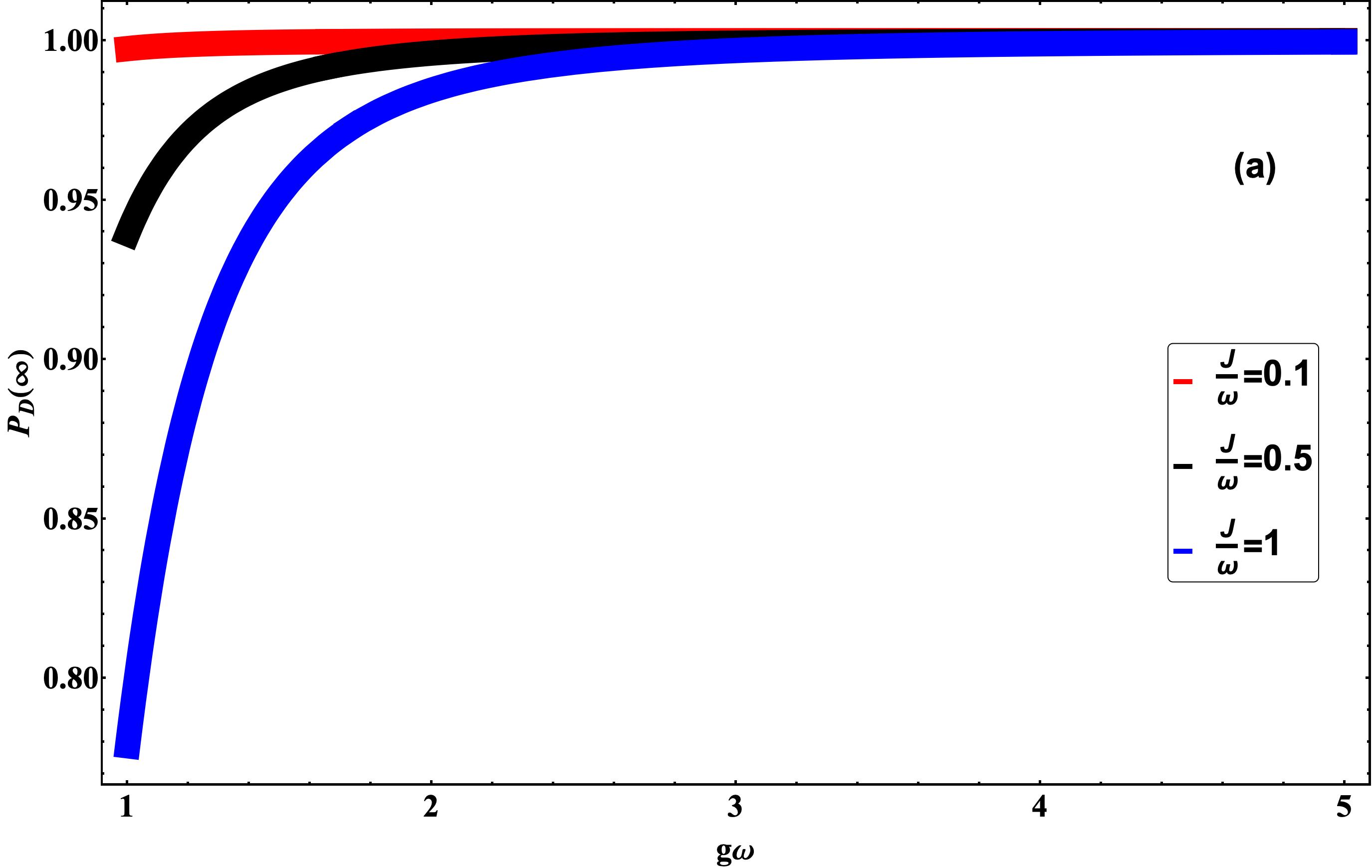}\hspace{2cm}
	\includegraphics[width=5.8cm,height=5cm]{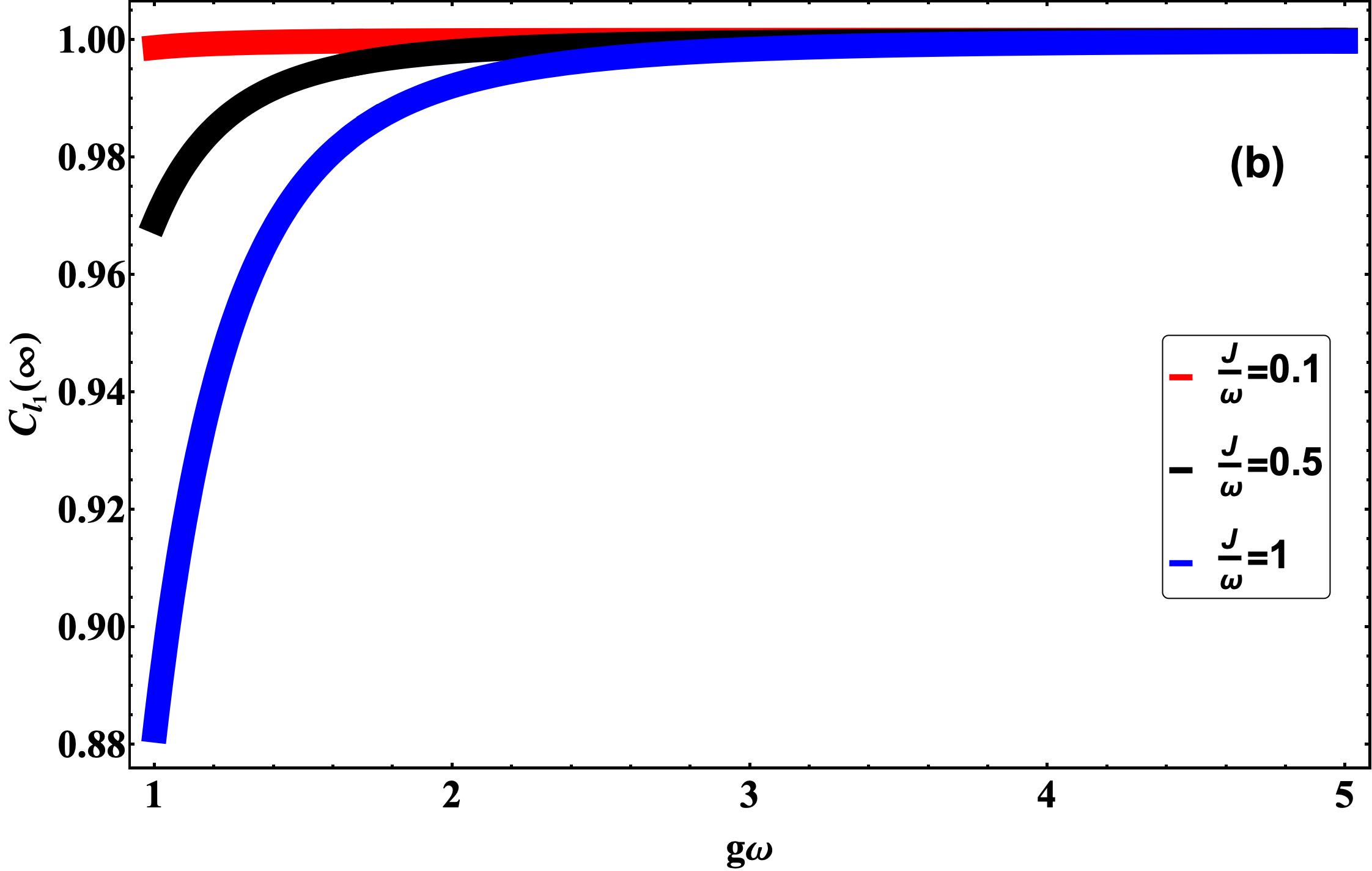}
	\caption{Here we plot (a) long time limit of population difference $P_D(\infty)$  (b) coherence $\mathcal{C}_{l_1}(\infty)$ with respect to $g \omega$ for various values of $\frac{J}{\omega}$. We observe that $P_D$ saturates at the maximum value due to coherent-incoherent transition in the model. The same is true for  $\mathcal{C}_{l_1}$. }
	\label{Large}
\end{figure}

	{In order to calculate the long time limit of the decoherence functions to recover the results of equation \ref{masi}, we have two parameters to control. First is the time parameter and other is the bare coupling $g\omega$.  First take the $t\rightarrow\infty$ limit at some fixed couplings and then  let $g\omega$ have large values. Now to take $t\rightarrow \infty$, we utilize a simple mathematical trick:
}
\begin{eqnarray}
		\underset{\eta\rightarrow0^+}{lim}\int_0^{\infty} dz \frac{e^{-i(a-i \eta)z}}{a-i\eta} =\frac{i}{a^2},
	\end{eqnarray}
	we can write $\int_0^{\infty}\frac{\sin n\omega t}{n\omega} =\frac{1}{n^2\omega^2}$ and $\int_0^{\infty} \frac{\cos n \omega t}{n \omega}=0$, therefore,
	$ \Gamma_{odd (even)}=4 \tilde{J}^2 \underset{n=odd (even)}{\sum} \frac{(4g^2\omega^2)^n}{n!} \frac{1}{n^2 \omega^2} $. Let $G=\sum_{n=odd}  \frac{(4g^2\omega^2)^n}{n!} \frac{1}{n^2 \omega^2} $. Differentiating $G$ twice with respect to $g^2$, we observe that $G$ satisfies the following differential equation for large $g$:
	\begin{eqnarray}
		g^4\frac{d^2 G}{dg^2} + g^2 \frac{dG}{dg^2}-\frac{e^{4g^2\omega^2}}{2\omega^2}=0.
	\end{eqnarray}
	In the large $g$ limit, this equation can be solved to give  the solution $G\sim \frac{e^{4g^2\omega^2}}{ 32g^4\omega^6} $. Therefore, in the long time limit we can write
	$\Gamma_{odd}= 4 \tilde{J}^2 \frac{e^{4g^2\omega^2}}{ 32g^4\omega^6}= \frac{J^2}{8 g^4 \omega^6}=\Gamma_{even} $. Similarly we can find the long time limit of $\Gamma(t) \sim \frac{J^2}{4g^2\omega^2} $. 
Therefore, we write
{
	\begin{eqnarray}
	\rho_{00}(t\rightarrow \infty) &=& \rho_{00}(0)\frac{1+ e^{-\Gamma(t\rightarrow\infty)}}{2} + \rho_{11}(0)\frac{1- e^{-\Gamma(t\rightarrow\infty)}}{2} \nonumber\\
	&=& \rho_{00}(0)\frac{1+ e^{-\frac{J^2}{4g^2\omega^2}}}{2} + \rho_{11}(0)\frac{1- e^{-\frac{J^2}{4g^2\omega^2} }}{2} \nonumber\\
	&\rightarrow& \rho_{00}(0) ~~~~~~{\rm for~~ g\omega >>1}.
\end{eqnarray}	
Similarly, we can find  same limit for the other elements of the density matrix and thus $ \underset{g\omega>>1}{\rm Limit}~ \underset{t\rightarrow\infty}{\rm Limit} \rho_S(t) =\rho_S(0)$ as obtained in the Markovian limit. These results are consistent with the plots drawn for $P_D(t\rightarrow \infty)$ and $C_{l_1}(\infty)$ in figure \ref{Large}. } We have,  $P_D(t\rightarrow \infty) \sim P_D(0) e^{-\frac{J^2}{4g^4\omega^6} }$. Also $\rho_{01}(t\rightarrow\infty) \sim \rho_{01}(0) e^{-\frac{J^2}{8 g^4 \omega^6}}$. This implies $C_{l_1}(\infty)=C_{l_1}(0) e^{-\frac{J^2}{8 g^4 \omega^6}}$. We plot $C_{l_1}(\infty)$ and $P_D (\infty)$ with respect to $g \omega$ in figure \ref{Large}(a) $\&$ \ref{Large}(b). We observe that there is finite amount of coherence in the system at large times.
.

	\section{Conclusions}
	In conclusion, we have studied spin boson model where a given qubit is  modeled by spin-$\frac{1}{2}$ particle that  is coupled strongly to a single bosonic mode. Next, we apply Lang-Firsov transformation to reduce the coupling strength in the polaron frame, where the qubit is dressed by phonons-known as polaron. 
	In the large polaron theory, where tunneling rates are significantly reduced, the system is expected to exhibit Markovian dynamics. We demonstrated this by directly extending the time integration limits from finite to infinite values in the time-convolutionless (TCL) master equation. Surprisingly, this approximation yielded no decoherence, which contradicts the fact that qubits can still decohere in a Markovian manner.
	To explore non-Markovian behaviour, we restricted our analysis to finite time evolution and employed coherence measures based on the $l_1$-norm. Our findings indicate that the system exhibits finite non-Markovian effects within a specific range of couplings, beyond which non-Markovianity vanishes. {
		 These effects are attributed to the resolution of various intermediate processes occurring in the second order perturbation. 
	
	It is important to note that all calculations are done zero temperature although the we derived master equation for finite temperature case as well. There are several works like \cite{FTM1,FTM2} dealing with finite temperature effects. A well mechanism based on the resonance effect has been proposed. There are mainly two reason to stick to zero temperature case: (1) the proliferation of phonon modes with different energies and (2) the perturbative analysis of these modes in a controlled fashion. The work in the present paper is mainly for single bath mode and its finite temperature generalization keeping all bath modes of same energy is a restricted condition. Therefore, we take up this problem in a different work. }

	\appendix
	\section{}
	
	In this appendix, we give detailed calculation of the terms involved in mater equation \ref{mas}. 
The master equation \ref{mas} can be written as
\begin{eqnarray}
	\label{mas1}
	\dot{\rho}^I_S(t) & =& -\int_0^{t} d\tau \big[{\rm Tr_B}[\mathcal{H}^I(t)\mathcal{H}^I(\tau)\rho^I_S (t)\otimes \rho_B] -
	{\rm Tr_B}[\mathcal{H}^I(t)\rho^I_S (t) \otimes \rho_B \mathcal{H}^I(\tau)] \big. \nonumber \\
	&&~~~~~~~~~~\big. - {\rm Tr_B}[\mathcal{H}^I(\tau)\rho^I_S (t)\otimes\rho_B \mathcal{H}^I(t)]
	+ {\rm Tr_B}[\rho^I_S (t) \otimes \rho_B \mathcal{H}^I(\tau) \mathcal{H}^I(t)]
	\big].
\end{eqnarray} 
It is sufficient to calculate the first term in the above equation \ref{mas1}, others are then straight forward to calculate. In the anti-adiabatic approximation equation \ref{ad}, the interaction Hamiltonian is given as $	\mathcal{H}^I(t) = \tilde{J}[ \sigma^{-}\mathcal{F}^{\dagger}(t)+ \sigma^+ \mathcal{F}(t)]$, therefore, we write
\begin{eqnarray}
	{\rm Tr_B}[\mathcal{H}^I(t)\mathcal{H}^I(\tau)\rho^I_S (t)\rho_B]= \tilde{J}^2  [\sigma^{-} \sigma^+\langle \mathcal{F}^{\dagger}(t)\mathcal{F}(\tau)\rangle_B + \sigma^+ \sigma^-\langle \mathcal{F}(t)\mathcal{F}^{\dagger}(\tau)\rangle_B],
\end{eqnarray}
where we have used $\sigma_+^{2}=0=\sigma_-^{2}$ and the $\langle....\rangle_B$ represents average with respect to bath. Assuming bath density matrix of the form $\rho_B=\frac{e^{-\beta \mathcal{H}_B}}{Z_B}$,
with $Z_B$ as the partition function of the bath, we write
\begin{eqnarray}
	\langle \mathcal{F}^{\dagger}(t)\mathcal{F}(\tau)\rangle_B&=& \frac{1}{Z_B} \sum_n e^{-\beta \omega_n} \bra{n_{ph}}(e^{-2g\omega b^{\dagger}e^{i\omega t}}e^{2g\omega be^{-i\omega t}}-1)(e^{2g\omega b^{\dagger}e^{i\omega \tau}}e^{-2g\omega b e^{-i\omega \tau}}-1)\ket{n_{ph}} \nonumber \\
	&=& 1+ \frac{1}{Z_B} \sum_n e^{-\beta \omega_n} \bra{n_{ph}}e^{-2g\omega b^{\dagger}e^{i\omega t}}e^{2g\omega b e^{-i\omega t}} e^{2g\omega b^{\dagger}e^{i\omega \tau}}e^{-2g\omega b e^{-i\omega \tau}}\ket{n_{ph}} \nonumber \\
	&& - \frac{1}{Z_B} \sum_n e^{-\beta \omega_n}  \bra{n_{ph}}e^{-2g\omega b^{\dagger}e^{i\omega t}}e^{2g\omega b e^{-i\omega t}}\ket{n_{ph}} 
	-\frac{1}{Z_B} \sum_n e^{-\beta \omega_n}  \bra{n_{ph}}e^{2g\omega b^{\dagger}e^{i\omega t}}e^{-2g\omega b e^{-i\omega t}}\ket{n_{ph}}.
	\label{ave}
\end{eqnarray}
Next, we observe that
\begin{eqnarray}
	e^{2g\omega b e^{-i\omega t}}\ket{n_{ph}}&=&\sum_{l=0}^{\infty}\frac{(2g\omega e^{-i\omega t})^l}{l!}b^{l}\ket{n_{ph}}\nonumber\\
	&=&\sum_{l=0}^{\infty}\frac{(2g\omega e^{-i\omega t})^l}{l!}\sqrt{\frac{n!}{(n-l)!}}\ket{n_{ph}-l}\nonumber\\
	\implies	\bra{n_{ph}}e^{-2g\omega b^{\dagger}e^{i\omega t}}e^{2g\omega b^{\dagger}e^{-i\omega t}}\ket{n_{ph}}
	&=&\sum_{l=0}^{\infty}\frac{[-(2g\omega)^2]^l}{l!}\frac{n!}{(n-l)!}=L_n[(2g\omega)^2].
\end{eqnarray}
 Here, we have used $b\ket{n_{ph}} =\sqrt{n}\ket{n_{ph}-1}$, and  $L_n[x^2]$ is the associated  Laguerre polynomial. The partition function of the bath is $Z_B= \sum_n e^{-\beta \omega_n}= \frac{1}{1-e^{-\beta \omega}}=\frac{1}{1-z}$, with $z=e^{-\beta \omega}$. Therefore, we write
\begin{eqnarray}
	\frac{1}{Z_B}\sum_{n}e^{-\beta \omega_{n}}\bra{n_{ph}}e^{-2g\omega b^\dagger e^{i\omega t}}e^{2g\omega b e^{-i\omega t}}\ket{n_{ph}}	
	&=& (1-z)\sum_{n}e^{-\beta \omega_n} L_n[(2g\omega)^2] =e^{(2g\omega)^2\frac{z}{z-1}}=e^{-(2g\omega)^2 N_0}
\end{eqnarray}
where we have used the fact $(1-z)\sum z^n L_n[x^2]= e^{x^2\frac{z}{z-1}}$ and $\frac{z}{z-1}=-N_0$, where $N_0=(1-e^{-\beta \omega})^{-1}$ is the Bose occupation number. Similarly we can evaluate the other terms in above equation \ref{ave}, thus we write 
\begin{eqnarray}
	\langle \mathcal{F}^{\dagger}(t)\mathcal{F}(\tau)\rangle_B&=& e^{4g^2 \omega^2 e^{-i\omega (t-\tau)}} e^{-8N_0g^2\omega^2 (1-\cos \omega (t-\tau))}-2e^{-4g^2\omega^2 N_0} +1 \\
	&\rightarrow& e^{4g^2 \omega^2 e^{-i\omega (t-\tau)}}-1~~~~~{\rm as~} \beta \rightarrow \infty
\end{eqnarray}
Similarly, we can write all the bath correlation functions that arise in evaluating the different terms in equation \ref{mas1}, as below:
\begin{eqnarray}
	\langle \mathcal{F}^{\dagger}(t)\mathcal{F}^\dagger(\tau)\rangle_B&=& \langle \mathcal{F}(t)\mathcal{F}(\tau)\rangle_B= e^{-4g^2 \omega^2 e^{-i\omega (t-\tau)}} e^{-8N_0g^2\omega^2 (1+\cos \omega (t-\tau))}-2e^{-4g^2\omega^2 N_0} +1 \\
	\langle \mathcal{F}^{\dagger}(t)\mathcal{F}(\tau)\rangle_B&=&\langle \mathcal{F}(t)\mathcal{F}^\dagger(\tau)\rangle_B= e^{4g^2 \omega^2 e^{-i\omega (t-\tau)}} e^{-8N_0g^2\omega^2 (1-\cos \omega (t-\tau))}-2e^{-4g^2\omega^2 N_0} +1.
\end{eqnarray}

Similarly, we evaluate the other terms in the double commutator which finally yields  the master equation given in the main text.

\end{document}